\documentclass{iopart}
\def\<{{<}} 
\def\>{{>}}

\usepackage{graphicx}
\input{epsf}
\usepackage{amssymb}
\begin{document}

\title{Reduced Google matrix}

\author{K.M.Frahm and D.L.Shepelyansky}
\address{Laboratoire de Physique Th\'eorique, IRSAMC, CNRS and
Universit\'e de Toulouse, UPS, 31062 Toulouse, France}

\begin{abstract}
Using parallels with the quantum scattering theory, 
developed for processes in nuclear and mesoscopic physics
and quantum chaos, we construct a reduced Google matrix $G_R$
which describes the properties and interactions of a certain subset
of selected nodes belonging to a much larger directed network.
The matrix  $G_R$ takes into account effective interactions between 
subset nodes by all their indirect links via the whole network.
We argue that this approach gives new possibilities to 
analyze effective interactions in a group of nodes embedded in
a large directed networks. Possible efficient numerical methods for 
the practical computation of $G_R$ are also described.
\end{abstract}




\vskip 0.3cm
{\bf Dated:} February 7, 2016


\submitto{\JPA}
\maketitle

\section{Introduction}

At present the concept of Markov chains 
finds impressive applications in descriptions of 
directed networks including the World Wide Web (WWW)
\cite{brin,googlebook}, citation networks \cite{redner},
Perron-Frobenius operators of dynamical systems \cite{mbrin},
software architecture \cite{linux}, Ulam networks of chaotic maps,
world trade flows, network of Wikipedia articles and many other
networks \cite{rmp}. Such directed networks are well described
by the Google matrix usually presented in the form
\begin{equation}
\label{eq_gmatrix}
G_{ij}=\alpha S_{ij}+(1-\alpha)/N \; ,
\end{equation}
where 
$S_{ij}$ describes Markov transitions on the network 
typically given by the inverse number of outgoing links from 
the node $j$ in presence of a link $j\to i$ or 0 in absence of such a link. 
In case of total absence of outgoing links from the node $j$ one 
replaces $S_{ij}=1/N$ for all values of $i$, i.~e. for the full column $j$ 
\cite{brin,googlebook}. 
A random surfer follows with probability $\alpha$, called damping factor, 
the dynamics fixed by the Markov 
transitions $S_{ij}$ and with the complementary probability $(1-\alpha)$ 
he jumps with uniform probability to any node of the $N$ nodes 
of the network. The elements of $G$ are non-negative and the sum of elements 
in each column is equal to unity corresponding to probability conservation.
As a result the product of two different Google matrices is also
a Google matrix, respecting these two properties. 

The eigenvalues $\lambda_i$ and right eigenvectors $\psi_i(j)$ of $G$
are defined by 
\begin{equation}
\label{eq_gmatrix2}
\sum_{j'} G_{jj'} \psi_i(j')=\lambda_i \psi_i(j) \; .
\end{equation}
The eigenvector at maximal $\lambda=1$ is called the PageRank vector.
It has only non-negative elements and, after normalizing its sum to unity, 
$P(j)$ has the meaning of the probability to find a random surfer 
on a given node $j$ in the stationary long time limit of the Markov process. 
Left eigenvectors are biorthogonal to right eigenvectors of different 
eigenvalues. 
The left eigenvector for $\lambda=1$ has identical (unit) entries 
due to the column sum normalization of $G$. 
One can show that the damping factor, when replacing $S$ by 
$G$ according to (\ref{eq_gmatrix}), only affects 
the PageRank vector (or other eigenvectors for $\lambda=1$ of $S$ in case 
of a degeneracy) while other eigenvectors are independent of $\alpha$ 
due to their orthogonality
to the left unit eigenvector for $\lambda=1$ \cite{googlebook} but 
their (complex) eigenvalues are reduced by a factor $\alpha$ 
when replacing $S$ by $G$. 
In the following we use the notations 
$\psi_L^T$  and $\psi_R$
for left and right eigenvectors  respectively
(here $T$ means  vector transposition).

In many real networks the number of nonzero elements
in a column of $S$ is significantly smaller then the whole matrix size $N$
that allows to find efficiently the PageRank vector by 
the PageRank algorithm of power iterations. Also
a certain number of largest eigenvalues (in modulus) and related eigenvectors
can be efficiently computed by the Arnoldi algorithm
(see e.g. \cite{arnoldibook,golub,ulamfrahm}).

At present directed networks of real systems can be very large
(e.g.  $4.5$ millions for the English Wikipedia edition in 2013 \cite{rmp} or
$3.5$ billion web pages for a publicly accessible web
crawl that was gathered by the Common Crawl Foundation in 2012
\cite{vigna}). In certain cases one may be interested in the 
particular interactions among a small reduced subset of $N_r$ nodes 
with $N_r \ll N$ instead of the interactions of the entire network. 
However, the interactions between these $N_r$ nodes should be 
correctly determined taking into account that there are many 
indirect links between the $N_r$ nodes via all other 
$N_s=N-N_r$ nodes of the network.
This leads to the problem of the reduced Google matrix $G_R$
with $N_r$ nodes which describes the interactions of
a subset of $N_r$ nodes.

In a certain sense we can trace parallels with 
the problem of quantum scattering
appearing in nuclear and mesoscopic physics 
(see e.g. \cite{sokolov1989,sokolov1992,beenakker,guhr,jalabert})
and quantum chaotic scattering (see e.g. \cite{gaspard}). 
Indeed, in the scattering problem there are effective
interactions between open channels to localized basis states in a well 
confined scattering domain where a particle can spend a certain time 
before its escape to open channels.
Having this analogy in mind we construct the reduced
Google matrix $G_R$ which describes interactions 
between selected $N_r$ nodes
and satisfies the standard requirements of the Google matrix.
This construction is described in the next Section 2 
and the discussion of the results
is given in Section 3.

\section{Determination of reduced Google matrix}

Let $G$ be a typical Google matrix of Perron-Frobenius type for 
a network with $N$ nodes such that $G_{ij}\ge 0$ and the 
column sum normalization $\sum_{i=1}^N G_{ij}=1$ are verified. 
We consider a sub-network 
with $N_r<N$ nodes, called ``reduced network''. In this case we can write 
$G$ in a block form~:
\begin{equation}
\label{eq_Gblock}
G=\left(\begin{array}{cc}
G_{rr} & G_{rs} \\
G_{sr} & G_{ss} \\
\end{array}\right)
\end{equation}
where the index ``$r$'' refers to the nodes of the reduced network and 
``$s$'' to the other $N_s=N-N_r$ nodes which form a complementary 
network which we will call ``scattering network''. 

Let us introduce the PageRank vector of the full network
\begin{equation}
\label{eq_Pageank0}
P=\left(\begin{array}{c}
P_r  \\
P_s  \\
\end{array}\right)
\end{equation}
which satisfies the equation $G\,P=P$ 
or in other words $P$ is the right eigenvector of $G$ for the 
unit eigenvalue. This eigenvalue equation reads in block notations:
\begin{eqnarray}
\label{eq_Pagerank1}
({\bf 1}-G_{rr})\,P_r-G_{rs}\,P_s&=&0,\\
\label{eq_Pagerank2}
-G_{sr}\,P_r+({\bf 1}-G_{ss})\,P_s&=&0.
\end{eqnarray}
Here ${\bf 1}$ is a unit diagonal matrix of corresponding size $N_r$
or $N_s$.
Assuming that the matrix ${\bf 1}-G_{ss}$ is not singular, i.e. all 
eigenvalues $G_{ss}$ are strictly smaller than unity (in modulus), we obtain 
from (\ref{eq_Pagerank2}) that $P_s=({\bf 1}-G_{ss})^{-1} G_{sr}\,P_r$
which gives together with (\ref{eq_Pagerank1}):
\begin{equation}
\label{eq_Geff1}
G_{\rm R}P_r=P_r\quad,\quad
G_{\rm R}=G_{rr}+G_{rs}({\bf 1}-G_{ss})^{-1} G_{sr}
\end{equation}
where the matrix $G_{\rm R}$ of size $N_r\times N_r$, defined for the 
reduced network, can be viewed as an effective reduced Google matrix. 
In this expression the contribution of $G_{rr}$ accounts for direct links 
in the reduced network and the second term with the matrix inverse 
corresponds to all contributions of indirect links of arbitrary order. 
We note that in mesocopic scattering problems one typically uses 
an expression of the scattering matrix which has a similar structure 
where the scattering channels correspond to the reduced network and 
the states inside the scattering domain to the scattering network 
\cite{beenakker}. 

The matrix elements of $G_R$ are non-negative since the matrix 
inverse in (\ref{eq_Geff1}) can be expanded as:
\begin{equation}
\label{eq_inverse_expand}
({\bf 1}-G_{ss})^{-1}=\sum_{l=0}^\infty G_{ss}^{\,l} \;\; .
\end{equation}
In (\ref{eq_inverse_expand}) 
the integer $l$ represents the order of indirect links, i.~e. the number 
of indirect links which are used to connect indirectly two nodes of the 
reduced network. The matrix inverse corresponds to an exact resummation 
of all orders of indirect links. 
According to (\ref{eq_inverse_expand}) 
the matrix  $({\bf 1}-G_{ss})^{-1}$ and therefore also $G_{\rm R}$
have non-negative matrix elements. 
It remains to show that $G_{\rm R}$ also fulfills the condition 
of column sum  normalization being unity. 
For this let us denote by $E^T=(1,\,\ldots,\,1)$ 
the line vector of size $N$ with unit entries and by $E_r^T$ (or $E_s^T$) 
the corresponding vectors for the reduced (or scattering) network 
with $N_r$ (or $N_s$) unit entries such that $E^T=(E_r^T,\,E_s^T)$. 
The column sum normalization for the full Google matrix $G$ implies that 
$E^T G=E^T$, i.~e. $E^T$ is the left eigenvector of $G$ with eigenvalue 
$1$. This equation becomes in block notation:
\begin{eqnarray}
\label{eq_leftE1}
E_r^T({\bf 1}-G_{rr})-E_s^T G_{sr}&=&0,\\
\label{eq_leftE2}
-E_r^T G_{rs}+E_s^T({\bf 1}- G_{ss})&=&0.
\end{eqnarray}
From (\ref{eq_leftE2}) we find that $E_s^T=E_r^T G_{rs}({\bf 1}- G_{ss})^{-1}$
which implies together with (\ref{eq_leftE1}) that 
$E_r^T G_{\rm R}=E_r^{T}$ using $G_{\rm R}$ as 
in (\ref{eq_Geff1}). This shows that the column sum normalization condition is 
indeed verified for $G_{\rm R}$ justifying that this matrix is indeed 
an effective Google matrix for the reduced network. 

The question arises how to evaluate practically the expression 
(\ref{eq_Geff1}) of $G_{\rm R}$ for a particular sparse and quite large 
network with a typical situation
when $N_r\sim 10^2$-$10^3$ is small compared to $N$ and $N_s \approx N\gg N_r$. 
If $N_s$ is too large (e.~g. $N_s\sim 10^5$) a direct naive evaluation 
of the matrix inverse $({\bf 1}- G_{ss})^{-1}$ in (\ref{eq_Geff1}) 
by Gauss algorithm is not possible. In this case we can try the 
expansion (\ref{eq_inverse_expand}) provided it converges sufficiently 
fast with a modest number of terms. However, this is most likely not the 
case for typical applications. 

Let us consider the situation 
where the full Google matrix has a well defined gap between the leading 
unit eigenvalue and the second largest eigenvalue (in modulus). For example 
if $G$ is defined using a damping factor $\alpha$ in the standard way, 
as in (\ref{eq_gmatrix}), the 
gap is at least $1-\alpha$ which is $0.15$ for the standard choice 
$\alpha=0.85$ \cite{googlebook}. 
For such a situation we expect that the matrix $G_{ss}$ 
has a leading real eigenvalue close to unity 
(but still different from unity so
that ${\bf 1}-G_{ss}$ is not singular) 
while the other eigenvalues are clearly 
below this leading eigenvalue with a gap comparable to the gap of 
the full Google matrix $G$. In order to evaluate the expansion 
(\ref{eq_inverse_expand}) efficiently, we need to take out analytically 
the contribution of the leading eigenvalue close to unity which is 
responsible for the slow convergence. 

In the following, 
we denote by $\lambda_c$ this leading eigenvalue and by $\psi_R$ 
($\psi_L^T$) the corresponding right (left) eigenvector such that 
$G_{ss}\psi_R=\lambda_c\psi_R$ (or $\psi_L^T G_{ss}=\lambda_c\psi_L^T$). 
Both left and right eigenvectors as well as $\lambda_c$ can be efficiently 
computed by the power iteration method in a similar way as the standard 
PageRank method. We note that one can easily  show that 
$\lambda_c$ must be real and that both left/right eigenvectors can be chosen 
with positive elements. Concerning the normalization for $\psi_R$ we 
choose $E_s^T\psi_R=1$ and for $\psi_L$ we choose $\psi_L^T\psi_R=1$. 
It is well known (and easy to show) that $\psi_L^T$ is orthogonal to all other 
right eigenvectors (and $\psi_R$ is orthogonal to all other 
left eigenvectors) of $G_{ss}$ with eigenvalues different from $\lambda_c$. 
We introduce the operator ${\cal P}_c=\psi_R\psi_L^T$ which is the 
projector onto the eigenspace of $\lambda_c$ and we denote by 
${\cal Q}_c={\bf 1}-{\cal P}_c$ the complementary projector. 
One verifies directly that both projectors commute with the matrix $G_{ss}$ 
and in particular ${\cal P}_c G_{ss}=G_{ss}{\cal P}_c=\lambda_c{\cal P}_c$. 
Therefore we can write:
\begin{eqnarray}
\label{eq_inverse_project1}
({\bf 1}-G_{ss})^{-1}&=&({\cal P}_c+{\cal Q}_c)({\bf 1}-G_{ss})^{-1}
({\cal P}_c+{\cal Q}_c)\\
\label{eq_inverse_project2}
&=&{\cal P}_c\frac{1}{1-\lambda_c}+
{\cal Q}_c({\bf 1}-G_{ss})^{-1}{\cal Q}_c\\
\label{eq_inverse_project3}
&=&{\cal P}_c\frac{1}{1-\lambda_c}+
({\bf 1}-\bar G_{ss})^{-1}{\cal Q}_c\\
\label{eq_inverse_project4}
&=&{\cal P}_c\frac{1}{1-\lambda_c}+
{\cal Q}_c \sum_{l=0}^\infty \bar G_{ss}^{\,l}
\end{eqnarray}
with $\bar G_{ss}={\cal Q}_c G_{ss}{\cal Q}_c$ and using the 
standard identity ${\cal P}_c{\cal Q}_c=0$ for complementary 
projectors. 
The expansion in (\ref{eq_inverse_project4}) has the advantage that 
it converges rapidly since $\bar G_{ss}^{\,l}\sim |\lambda_{c,2}|^l$ 
with $\lambda_{c,2}$ being the second largest eigenvalue which is 
significantly 
lower than unity (e.~g. $|\lambda_{c,2}|\approx \alpha=0.85$ for the case 
with a damping factor). The first contribution due to the leading 
eigenvalue $\lambda_c$ close to unity is taken out analytically once 
the left and right eigenvectors, and therefore also the projector 
${\cal P}_c$, are known. 
The combination of (\ref{eq_Geff1}) and (\ref{eq_inverse_project4}) 
provides an explicit algorithm feasible for a numerical implementation 
for the case of modest values of $N_r$, large values of $N_s$ and of course 
for sparse matrices $G$, $G_{ss}$ etc. 

The method can also be modified to take out analytically the contributions 
of several leading eigenvalues close to unity if the latter are sufficiently 
well separated (non-degenerate) such that these eigenvalues and left/right 
eigenvectors can be determined by the Arnoldi method (applied to $G_{ss}$). 
Then Eq. (\ref{eq_inverse_project4}) is modified as:
\begin{equation}
\label{eq_inverse_modified}
({\bf 1}-G_{ss})^{-1}=\sum_j {\cal P}_c^{(j)}\frac{1}{1-\lambda_{c,j}}+
{\cal Q}_c \sum_{l=0}^\infty \bar G_{ss}^{\,l}
\end{equation}
with ${\cal P}_c^{(j)}=\psi_R^{(j)}(\psi_L^{(j)})^T$ being the projector 
on the eigenspace of the eigenvalue $\lambda_{c,j}$ with right (left) 
eigenvector $\psi_R^{(j)}$ [or $(\psi_L^{(j)})^T$] obeying, after 
proper normalization, the bi-orthogonality 
identity $(\psi_L^{(j)})^T\psi_R^{(k)}=\delta_{jk}$ and 
with ${\cal Q}_c={\bf 1}-\sum_j {\cal P}_c^{(j)}$ 
being the total complementary 
projector. The expression (\ref{eq_inverse_modified}) is in principle 
also suitable for a numerical evaluation provided that the number of 
leading eigenvalues $\lambda_{c,j}$ is modest. 

We note that the numerical methods described in \cite{fgsjphysa}
allow to determine the eigenvalues $\lambda_c$ 
(and corresponding eigenvectors) 
which are exponentially close to unity
(e.g. $1-\lambda_c \sim 10^{-16}$) so that the expression 
(\ref{eq_inverse_modified}) can be efficiently computed numerically.

In the case when $N < 20000$ an exact diagonalization
of $G_{ss}$ can be done numerically and then
the presentation $G_{ss}= Q D_\lambda Q^{-1}$
allows to obtain the simple expression
$({\bf 1}-G_{ss})^{-1} = Q [{\bf 1}/({\bf 1}-D_\lambda)] Q^{-1}$.
Here $Q$ is the regular matrix formed by eigenvectors of $G_{ss}$ 
(in its columns) and $D_\lambda$ is the diagonal matrix of 
corresponding (complex) eigenvalues $\lambda$.

There is also an additional possibility to avoid the problem 
of slow convergence in $G_{\rm R}$ by a slight modification of the 
initial Google matrix to the form 
\begin{equation}
\label{eq_Gblock2}
G_{\rm mod}=
\left(\begin{array}{cc}
{\bf 1} & (1-\eta) U_{rs}\\
0 & \eta {\bf 1} \\
\end{array}\right) \times
\left(\begin{array}{cc}
G_{rr} & G_{rs} \\
G_{sr} & G_{ss} \\
\end{array}\right) \; .
\end{equation}
Here $0.5 \leq \eta <1$ is an additional damping factor,
$U_{rs}$ is a rectangular $N_r \times N_s$ matrix with non-negative 
elements and whose columns are sum normalized. A possible choice
is $U_{rs}=(1/N_r)E_r E_s^T$ with $E_r$ or $E_s$ as defined in 
the paragraph preceeding Eq. (\ref{eq_leftE1}) or more generally 
$U_{rs}=v_p E_s^T$ where $v_p$ is a sum normalized vector with $N_r$ 
non-negative entries and representing somehow a kind of preferential 
vector on the reduced network. 
Therefore the first matrix in the product of 
(\ref{eq_Gblock2}) belongs to the Google matrix class
(sum of non-negative elements in each column is equal to unity). 
Thus, the product of both matrices also belongs 
to the class of Google matrices
and hence $G_{\rm mod}$ is also a matrix of Google type.
Then for $G_{\rm mod}$, in analogy with
(\ref{eq_Geff1}), we obtained the modified reduced Google matrix
\begin{equation}
\label{eq_Geffmod}
{\hskip -2.0cm}
G_{\rm Rmod}=G_{rr}+(1-\eta)U_{rs} G_{sr} + 
\eta [G_{rs}+(1 -\eta)U_{rs} G_{ss}]({\bf 1}-\eta G_{ss})^{-1} G_{sr} \; .
\end{equation}
If $\eta$ is sufficiently smaller than unity, e.~g. $1-\eta\approx 0.1$-$0.2$,
then 
the geometric series expansion analogous to (\ref{eq_inverse_expand})
converges rapidly allowing for an efficient numerical computation even 
if $G_{ss}$ has a maximal eigenvalue close to unity. 
We note that a similar expansion has
been used for the ImpactRank in \cite{physrev} where the rapid numerical 
convergence allowed for an efficient computation. 

Finally we note that in a similar way it is possible to construct
the reduced matrix for the network of same $N$ nodes
with the inverted direction of links. This gives the 
Google matrix $G^*$ with the CheiRank eigenvector $P^*$
of $G^*$ at $\lambda=1$ \cite{linux,rmp}.
Then from $G^*$ using (\ref{eq_Geff1}) we obtain the reduced
matrix ${G_R}^*$.

\section{Discussion}

The obtained expression (\ref{eq_Geff1}) for the reduced Google matrix $G_R$
allows to analyze effective interactions between a selected
subset of nodes of a given large network.
We expect that this will allow to understand in a better way
hidden indirect dependencies existing between specific nodes in 
small subsets of large networks. The geometric series expansion 
of the propagator $({\bf 1}-G_{ss})^{-1}$ in (\ref{eq_inverse_expand})
is similar to the propagators appearing in the theory of quantum scattering
\cite{sokolov1992,beenakker,gaspard} corresponding to summation over all periods of 
particle motion inside the confined scattering domain. 
In our case $N_s$ nodes correspond to localized basis states 
in the scattering domain while the subset of $N_r$ nodes in the 
reduced network describes interactions (scattering)
between open channels.
We think that such an analogy will find further useful applications.
The reduced Google matrix should allow to study 
effective interactions between a small group of friends.
For a group of three, four friends it would be interesting to compare
results from real networks with the known results 
for the ensemble of random orthostochastic matrices \cite{karol}.
It would be also interesting to analyze the properties of $G_R$ for 
a class of random RPFM matrices considered in \cite{physrev}
(see Fig.16 there).

Thus we expect that the description of specific subsets of directed
networks with the help of the reduced Google matrix will find many 
interesting applications.

\normalsize

\end{document}